\documentclass[12pt]{iopart}
\usepackage{iopams,setstack}
\usepackage[T1]{fontenc}
\usepackage{graphicx}
\newcommand{\lpt}      {\ensuremath{\left.}}
\newcommand{\rpt}      {\ensuremath{\right.}}
\newcommand{\lp}       {\ensuremath{\left(}}
\newcommand{\rp}       {\ensuremath{\right)}}
\newcommand{\lb}       {\ensuremath{\left[}}
\newcommand{\rb}       {\ensuremath{\right]}}
\newcommand{\lv}       {\ensuremath{\left|}}
\newcommand{\rv}       {\ensuremath{\right|}}
\newcommand{\lk}       {\ensuremath{\left\{}}
\newcommand{\rk}       {\ensuremath{\right\}}}
\newcommand{\lr}       {\ensuremath{\left\langle}}
\newcommand{\rr}       {\ensuremath{\right\rangle}}
\newcommand{\be}       {\begin{equation}}
\newcommand{\ee}       {\end{equation}}
\newcommand{\nnb}      {\nonumber}

\newcommand{\abs}[1]   {\ensuremath{ \lv #1 \rv }}
\newcommand{\dfrac}    {\ensuremath{\displaystyle\frac}}
\newcommand{\avg}[1]   {\ensuremath{\overline{#1}}}
\newcommand{\rav}[1]   {\ensuremath{\left\langle#1\right\rangle}}
\renewcommand{\tr}[1]  {\ensuremath{\lr\lr #1 \rr\rr}}
\newcommand{\eav}[1]   {\ensuremath{\overbrace{#1}}}
\newcommand{\tav}[2][T]{\ensuremath{\left\langle#2\right\rangle_{#1}}}
\newcommand{\xsim}[1]  {\overset{#1 \gg 1}{\sim}}

\newcommand{\ket}[1]   {\ensuremath{\lv #1 \rr}}
\newcommand{\Ket}[2][E]{\ensuremath{\ket{#1_{#2}}}}
\newcommand{\bra}[1]   {\ensuremath{\lr #1 \rv}}
\newcommand{\Bra}[2][E]{\ensuremath{\bra{#1_{#2}}}}

\newcommand{\BOK}[3]   {\ensuremath{\lr #1 \rv #2 \lv #3 \rr}}
\newcommand{\KOK}[2]   {\ensuremath{\lr #1 \rv #2 \lv #1 \rr}}
\newcommand{\Op}[1][O] {\ensuremath{\mathcal{#1}}}

\newcommand{\DO}[2][_o] {\ensuremath{D{#1}\!\lp E_{#2} \rp}}

\begin{document}

\title{Thermalization in the Two-Body Random Ensemble}

\author{V. K. B. Kota$^1$, A. Rela\~no$^2$, J. Retamosa$^3$, Manan Vyas$^1$}
\address{$^1$ Physical Research Laboratory, Ahmedabad 380 009, India}

\address{$^2$ Instituto de Estructura de la Materia, IEM-CSIC Serrano, 123, 28006-Madrid. Spain}

\address{$^3$ Grupo de F\'{\i}sica Nuclear, Departamento de F\'{\i}sica  At\'omica, Molecular y Nuclear, Universidad Complutense de Madrid, E-28040 Madrid, Spain}

\begin{abstract}
Using the ergodicity principle for the expectation values of several
types of observables, we investigate the thermalization process in
isolated fermionic systems. These are described by the two-body random
ensemble, which is a paradigmatic model to study quantum chaos and
specially the dynamical transition from integrability to chaos. By
means of exact diagonalizations we analyze the relevance of the
eigenstate thermalization hypothesis as well as the influence of other
factors, like the energy and structure of the initial state, or the
dimension of the Hilbert space. We also obtain analytical expressions
linking the degree of thermalization for a given observable with the
so-called number of principal components for transition strengths
originated at a given energy, with the dimensions of the whole Hilbert
space and microcanonical energy shell, and with the correlations
generated by the observable. As the strength of the residual
interaction is increased an order-to-chaos transition takes place, and
we show that the onset of Wigner spectral fluctuations, which is the
standard signature of chaos, is not sufficient to guarantee
thermalization in finite systems. When all the signatures of chaos are
fulfilled, including the quasi complete delocalization of
eigenfunctions, the eigenstate thermalization hypothesis is the
mechanism responsible for the thermalization of certain types of
observables, such as (linear combinations of) occupancies and strength
function operators. Our results also suggest that fully chaotic
systems will thermalize relative to most observables in the
thermodynamic limit.
\end{abstract}

\maketitle

\newpage

\section{Introduction}

The success of thermodynamics is based on the fact that the state of a macroscopic system behaves as
it were independent of the microscopic details of the system, and thus it can be determined from a
few universal laws.  In classical isolated systems this thermodynamic universality can be explained
satisfactorily if one assumes that the dynamics is ergodic and mixing. These properties define
classical deterministic chaos~\cite{Reichel:80} and lead to the equiprobability of equal-volume
regions of the available phase space, which is the starting point of the microcanonical ensemble
formulation. This implies that, on average, the state of the system is independent of the initial
conditions as well as independent of the measurement time. It is thus generally accepted that the
system equilibrates into a thermal state, described by the microcanonical ensemble, when its
dynamical regime is ergodic. On the other hand thermalization is expected to be inhibited for
integrable or quasi integrable systems.  In these systems (almost) every trajectory in phase space
lies in one of the so-called invariant tori, which foliate the phase space. Therefore, being
restricted the trajectories to these structures, the equiprobability equal-volume regions of the
available phase space does not hold and the system will not thermalize, at least in the sense
explained above.

In recent years the study of the equilibration and thermalization mechanisms in isolated quantum
systems has attracted a great interest partly because the non-equilibrium dynamics, after an
external perturbation is applied, has become experimentally accessible for ultra-cold quantum
gases~\cite{Jordens:08,Schneider:08} and electrons in solids~\cite{Perfetti:06}. The technology
makes it possible to induce sharp changes in the parameters controlling the system and then observe
the subsequent time evolution, which is essentially unitary because on short and intermediate time
scales the perturbed system is almost isolated from the environment. Thus, one can experimentally
study if an isolated system equilibrates after a sharp perturbation and in this case whether it
thermalizes or retains memory of the initial conditions.  Experimental studies of the
non-equilibrium dynamics in one-dimensional ultra-cold Bose gases have given, up to the moment,
contradictory results~\cite{Kinoshita:06,Hofferberth:07}

From a theoretical point of view, these questions have been addressed using different methods and
points of view. The results from classical mechanics can not be translated directly into quantum
mechanics. On one hand the concept of quantum integrability is not well defined, though it is
generally considered as synonym for exact solvability~\cite{Feng:95,Relano_Dukelsky:04}. On the
other hand the unitary time evolution of quantum states leaves no room for a dynamical regime
similar to deterministic chaos. Actually, the name quantum chaos stands for the different type of
signatures that certain quantum systems exhibit depending on the large time scale behavior of their
classical analogues~\cite{Stockmann:99,Haake:10}. These signatures appear on the statistical
behavior of eigenenergies and eigenfunctions and is not clear at all how they can influence the
evolution of the system on large-time scales.

The generally accepted assumption that integrable systems do not thermalize is corroborated in
several models~\cite{Igloi_Rieger:00,Cazalilla:06,Calabrese_Cardy:06,Segumpta_Powell:04,
  Chern_Levitov:06}, where the non-equilibrium regime extends over large time scales or where a
long-time steady state with non standard thermal properties is reached. The approach to thermal
equilibrium of generic systems has been studied by several authors. It has been proven that almost
any system in interaction with a large heat bath will equilibrate and
thermalize~\cite{Tasaki:98,Linden_Popescu:09}. For isolated systems it has been shown that
``typical'' Hamiltonian and observables will be in thermal equilibrium for most
times~\cite{Reimann:08,Goldstein_Lebowitz:10}, but it seems a very difficult task to decide whether
a specific Hamiltonian or observable belongs to this class or not. The thermalization of specific 1D
and 2D fermionic and bosonic systems has also been
studied~\cite{Kollath_Lauchli:07,Manmana_Wessel:07,Rigol_Dunjko:08,Rigol:09a,Rigol:09b}.
Surprisingly, thermalization was not obtained in all the cases.  Several reasons have been reported
to explain the lack of thermalization in these systems, but it seems that the so-called eigenstate
thermalization hypothesis (ETH) plays a fundamental role~\cite{Rigol:09a,Rigol:09b}. The ETH states
that thermalization occurs at the level of individual eigenstates~\cite{Deutsch:91,Srednicki:94},
whenever they satisfy Berry's conjecture on chaotic eigenfunctions~\cite{Berry:77a}, i.e., whenever
they behave as (quasi) random superpositions of the basis states.  For this and other reasons, the
role played in the viability of thermalization by quantum chaos in general, and by the properties of
chaotic wave functions in particular, has began to receive some attention~\cite{Santos_Rigol:10,
  Relano:10}. It seems that the number of principal components (NPC) or inverse participation ratio
(IPR), which keeps track of the progressive eigenstate delocalization through the integrability to
chaos transition, might be directly related to the deviations of the steady expectation values from
the corresponding statistical values~\cite{Olshanii_Yurovsky:09,Neuenhahn_Marquardt:10}

In order to get a deeper understanding of the role played by quantum chaos in these processes we
study the thermalization of isolated fermionic systems modeled by the simplest two-body random
ensemble~\cite{French:80}. Two-body random ensembles are paradigmatic models to study quantum chaos
and specially the dynamical transition from integrability to chaos. They have also been used in the
past, together with some related models, such as nuclear shell model and interacting spin models, to
perform different studies on thermalization. The thermalization criteria were based on the
equivalence between different definitions of entropy~\cite{Horoi:95,Kota_Sahu:02,Brody_Flores:81} and
temperature~\cite{Borgonovi:98}; representability of occupancies by Fermi-Dirac
distribution~\cite{Flambaum:96,Flambaum:97,Beneti:01} (Bose-Einstein distribution for bosons); and
calculation of expectation values using the canonical distribution~\cite{Flambaum:97}.  A brief
review of some of these studies can be found in Refs.~\cite{Kota:01,Gomez_Kar:11}. However, in the
present work, as well as in most recent papers the focus is put on the Ergodicity
principle~\cite{Reimann:08,Goldstein_Lebowitz:10} which is the cornerstone for thermalization, and
clearly more precise and general that the aforementioned criteria. We study the relevance of several
factors in the thermalization process like ETH, the dimension of the Hilbert space, the structure
and energy of the initial state and its proximity to the ground state. We also analyze the
importance of the degree of chaos as measured by the different chaos markers.

The rest of the paper is organized as follows. Sec. \ref{sec:model_transitions} briefly introduces
the embedded Gaussian ensembles of random matrices generated by two-body interactions, with emphasis
in the embedded Gaussian orthogonal ensemble, and some of the well established main features of the
order to chaos transition in this ensemble. However, the criteria for determining the transition
points as the two-body interaction strength is increased, are somewhat different from those used in
the past.  Sec. \ref{sec:thermalization} deals with the thermalization of embedded Gaussian
orthogonal ensembles. Here we give all the new results of the paper, where the transition points of
Sec \ref{ssec:order2chaos} play a central role. After introducing some basic definitions in
Sec. \ref{ssec:definitions}, Sec. \ref{ssec:NResults} reports the main numerical results linking
thermalization with the type of spectral fluctuations, the delocalization of the wave functions, the
structure and the proximity of the initial state to the ground state, or the dimension of the
Hilbert space.  In Sec. \ref{sec:AResults} we gather together some analytical results that
establish a connection between thermalization, relative to a given observable $\Op$, and the value
of the NPC for the transition strengths originated at a given eigenstate, or between thermalization
and the dimension of the whole Hilbert space, the dimension of the microcanonical energy shell and
the correlations generated by $\Op$. Finally, Sec. \ref{sec:Conclusions} contains the conclusions.

\section{EGOE(1+2) model: order-to-chaos transitions}
\label{sec:model_transitions}

\subsection{EGOE(1+2) model}
\label{ssec:model}

As  previously mentioned, we try to analyze the relationship between the order-to-chaos transition
and the thermalization of fermionic systems, which can be modeled by an appropriate two-body random
matrix ensemble. Here we briefly introduce these ensembles and their relation with quantum
chaos. Recent and comprehensive reviews can be found in references~\cite{Kota:01,Gomez_Kar:11}.

There is a clear relationship between the energy level fluctuation properties of a quantum system
and the large time scale behavior of its classical analogue.  The spectral fluctuations of a quantum
system whose classical analogue is fully integrable are well described by Poisson statistics, i.e.,
the spacings between successive energy levels are not correlated~\cite{Berry:77b}.  According to
Bohigas, Gianoni and Schmit~\cite{Bohigas:84}, the fluctuation properties of generic quantum
systems, which in the classical limit are ergodic, coincide with those of the Gaussian ensembles
(GE) of random matrices~\cite{Mehta:04}.  This statement, initially supported by many experimental
data and numerical calculations, has been finally proven for quantum systems with few degrees of
freedom, where the semiclassical approximation is valid~\cite{Heusler_Muller:07}.  The large time
scale behavior of the classical analogue also determines the properties of the wave
functions. Extensive reviews of later developments can be found in \cite{Haake:10,Gomez_Kar:11} and
references there in.

Quantum many-body systems like complex atoms and atomic nuclei are usually considered to be chaotic
if their spectral fluctuations are those of the Gaussian orthogonal ensemble (GOE), which is the
appropriate GE for systems with time-reversal invariance and rotational symmetry.  However, real
quantum systems are usually well described by real or effective one- plus two-body interactions in
the mean-field basis, whilst GE represent systems with multi-body interactions.  The embedded
Gaussian ensembles (EGE) of random matrices were introduced to tackle this problem, and to provide a
more realistic picture of many-body quantum systems. Moreover, in the present context EGE are
interesting because random interactions can illustrate the effects on thermalization caused by
generic interactions, which lead the system from integrability to chaos.

The EGE(1+2) ensembles consider $m$ fermions or bosons distributed in $n$ single-particle states
$\ket{k},\;\; k=1,2,\cdots,n$, interacting via the following Hamiltonian 

\be
\label{eq:H(1+2)}
H = \sum_k \varepsilon_k a^{\dagger}_{k}a_{k} + \lambda\sum_{k\leq l,p\leq q} \BOK{p q}{V}{k l} a^{\dagger}_{p}a^{\dagger}_{q}a_{l}a_{k},
\ee
where the single-particle energies $\varepsilon_k$ and the two-body matrix elements (properly
symmetrized or antisymmetrized) $\BOK{p q}{V}{k l}$ behave as independent Gaussian random variables.
In this expression $\lambda$ gives the strength of the two-body interaction, and $a^{\dagger}_{k}$
and $a_{k}$ create and destroy a fermion (or a boson) in the $k$th single-particle state. In
fermionic systems only the strict inequalities $k< l$ and $p < q$ are valid.  It has been
numerically shown that for EGE(2) ensembles ($\varepsilon_k=0$) the spectral fluctuations do agree
with those of GE, provided that energy scale is redefined appropriately. Introducing the notation
$\rav{\bullet} = \tr{\bullet}/ d$, where $\tr{\bullet}$ stands for the trace operation, and $d$ is
the dimension of the Hilbert space, the centroid and the energy span must be $\rav{E} =0$ and
$\rav{E^2}^{1/2}=1$, respectively.~\cite{Brody_Flores:81}

\subsection{The order-to-chaos transition in EGOE(1+2)}
\label{ssec:order2chaos}

One of the most significant aspects of EGOE(1+2) is that as $\lambda$ increases, starting from
$\lambda=0$, the system undergoes a transition from a regular to a chaotic regime that deeply
affects the state density, level fluctuations, and wave functions. This change of dynamical regime
is characterized by three chaos markers~\cite{Kota:01,Gomez_Kar:11,Vyas_Kota:10,Angom_Ghosh:04}.
There is a first marker $\lambda_c$ that signals the transition from Poisson to GOE spectral
fluctuations.  This transition occurs when the interaction strength $\lambda$ is of the order of the
spacing between the basis states that are directly coupled by the residual two-body interaction, a
result that came out of nuclear structure calculations by {\AA}berg \cite{Aberg:90,Aberg:92}. More
developments in determining the $\lambda_c$ marker are given for example
in~\cite{Jackod_Varga:02,Jacquod_Shepelyansky:97}. An important outcome of {\AA}berg criterion is
that $\lambda_c \propto 1/(m^2n)$ for EGOE(1+2), which is well verified in
~\cite{Jacquod_Shepelyansky:97} and in the results presented below.

As $\lambda$ increases further from $\lambda_c$, the structure of the eigenstates undergoes a deep
transformation. First the strength functions change from Breit-Wigner to Gaussian at a transition
point denoted by $\lambda_F$.  Beyond $\lambda_F$ we find a third chaos marker $\lambda_t$ which
defines the center of a region where different definitions of thermodynamic variables, such as
entropy, temperature, specific heat, etc., give the same results, as it occurs for infinite
systems. As far as the statistical entropy $S^{ther}$ and the Shanon entropy $S^{inf}$ are concerned
we can understand the meaning of $\lambda_t$ as follows. The former is proportional to the logarithm
of the state density, which in our case has essentially Gaussian form, though its mean and variance
depend on $\lambda$. At $\lambda=0$, where the eigenstates are fully localized in the mean-field
basis, information entropy is $S^{inf}=0$, whilst for sufficiently large values of the interaction
strength, the eigenstates are quite similar to those of GOE. In our case this occurs for
$\lambda\gtrsim 1$ and then $S^{inf} \approx \log(0.48d)$, except perhaps near the spectrum edges.
Changing from $\lambda=0$ to $\lambda=1$ the wave functions become more and more delocalized in the
mean-field basis, being this process faster in the middle than in the spectrum edges, and
$\lambda_t$ signals the precise value of the residual interaction strength in which the energy
dependence of $S^{ther}$ and $S^{inf}$ is quite similar. Thus, if we define an appropriate
``distance'' between the two entropies, it should have a minimum at $\lambda_t$.  Another property
is that $\lambda_t$ signals the duality point between $h(1)$ and $V(2)$ basis, i.e., the point where
the eigenstates become equally delocalized in the two
basis~\cite{Angom_Ghosh:04,Jackod_Varga:02}. Beyond $\lambda_t$ the eigenstates become quickly
similar to those of GOE systems, for which they are essentially Gaussian random superposition of the
basis states. This property resembles Berry's conjecture about the {\em ergodic} structure of
chaotic wave functions in phase space~\cite{Berry:77a}.  Nevertheless, a word of caution should be
included here. Since $S^{inf}$ is basis dependent, for $\lambda \gg \lambda_t$ it would be more
appropriate to define $S^{inf}$ in the $V(2)$ basis.

For our purposes the markers $\lambda_c$ and $\lambda_t$ are the most relevant. We detect the change
in the spectral fluctuations, and thus the position of $\lambda_c$, using the nearest-neighbor
spacing distribution, denoted $P(s)$. The spacings for generic integrable systems obey the Poisson
distribution, i.e, $P(s)= \exp{(-s)}$, while the Wigner surmise, $P(s)=\frac{\pi}{2} s \exp{\lp
  -\frac{\pi}{4} s^2\rp}$, provides a very good approximation for GOE-like
systems~\cite{Mehta:04}. The Brody distribution~\cite{Brody_Flores:81},

\be
P(s,\omega)= A_{\omega}(\omega+1) s^{\omega}\exp{(-A_{\omega} s^{\omega+1})}, 
\label{eq:brody}
\ee
where $\omega$ is usually called the Brody parameter and $A_{\omega}$ is a normalization constant,
it is used to asses how close the fluctuations are to the Poisson limit, corresponding to
$\omega=0$, or to the Wigner surmise with $\omega=1$~\cite{omegaGOE}. Although Eq. \eref{eq:brody}
is only a heuristic formula, and the Brody parameter has no definite meaning for Hamiltonian
systems, it has been employed in a variety of studies since its introduction. Very recently a
physical foundation of $\omega$ has been found by Sakhr and Nieminen in the context of self-similar
fractals~\cite{Sakhr_Nieminen:05}.  As the interaction strength in Eq. \eref{eq:H(1+2)} increases
from $\lambda=0$ to a sufficiently large value, the Brody parameter changes from $\omega=0$ to
$\omega \simeq 1$. In what follows, the position of the first chaos marker $\lambda_c$ is fixed by
the condition $\omega(\lambda) = 1/2$. Similar conditions can be found in the
literature~\cite{Jacquod_Shepelyansky:97, Berkovits_Avishai:98}.

To locate the value of the third marker $\lambda_t$ we consider the values of three different entropies:

\begin{itemize}
\item Thermodynamic entropy,  $S^{ther}_E = \log \rho(E) $, where $\rho(E)$ is the density of states.

\item Information entropy in the mean-field basis $S^{inf}_E  = -\sum_{k=1}^d\; \lv c_k^E \rv^2
\; \log\;\lv c_k^E \rv^2$, where the coefficients $c_k^E$ are the eigenstates components in the
mean-field basis.

\item Single-particle entropy $S^{sp}_E= -\sum\; \{ \lr n_i \rr^E \log \lp\lr n_i \rr^E\rp + \lp
  1-\lr n_i \rr^E\rp \log \lp 1-\lr n_i \rr^E \rp\}$, with $\lr n_i \rr^E$ the occupancy of the
  $i$th single-particle state at energy $E$.
\end{itemize}

Entropies formulas, valid for $\lambda > \lambda_F$, are given in~\cite{Kota_Sahu:02,Kota_Sahu:01}
for EGOE(1+2) ensembles. If we define the average ``distance'' between the three entropies as

\be
\Delta_s(\lambda) = \lk \displaystyle\int_{-\infty}^{\infty} \lb \lp R^{inf}_E-R^{ther}_E\rp^2 \rpt\rpt\!\!
                  + \lpt\lpt\lp R^{sp}_E-R^{ther}_E\rp^2\rb d\,E\rk^{1/2} \Big/ {\displaystyle\int_{-\infty}^{\infty} R^{ther}_E d\,E},
\ee
where $R^{\alpha}_E = \exp\lk S^{\alpha}_E-S_{max}^{\alpha}\rk$, the value of $\lambda_t$
corresponds to the minimum of $\Delta_s$ because this ensures that the values of the different
entropies will be very close to each other.

\begin{figure}[h]
\begin{center}
\includegraphics[width=0.35\textwidth,height=0.4\textwidth,angle=-90]{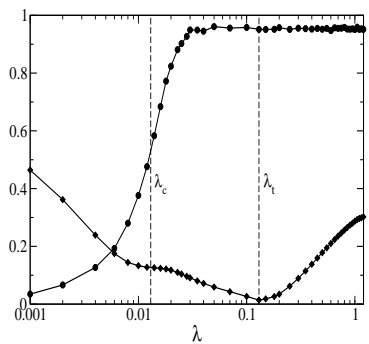} 
\vskip-3ex
\caption{Average values $\avg{\omega(\lambda)}$ (dots) and $\avg{\Delta_s(\lambda)}$ (diamonds) for
$\lambda\in[0,1]$, calculated using $60$ members of a EGOE(1+2) with $m=6$ and $n=16$. The vertical
dashed lines indicate the position of $\lambda_c$ and $\lambda_t$.}
\label{fig:AvgChaosMarkers_6x16_60mt}
\end{center}
\end{figure}

Fig. \ref{fig:AvgChaosMarkers_6x16_60mt} displays the ensemble averages $\avg{\omega(\lambda)}$ and
$\avg{\Delta_s(\lambda)}$ along $\lambda\in[0,1]$, for a $60$ member EGOE(1+2) with $m=6$ and
$n=16$. In order to enlarge the region where the order-to-chaos transition takes place we use a
logarithmic horizontal axis. The two vertical dashed lines indicate the respective positions of
$\lambda_c$ and $\lambda_t$. It can be seen that $\lambda_c \simeq 0.013$ and $\lambda_t \simeq
0.13$, values that are consistent with theoretical estimates
(see~\cite{Kota:01,Gomez_Kar:11,Vyas_Kota:10}). For $\lambda \simeq 0.03$ we obtain $\omega \simeq
0.96$, which is very close to the actual GOE result. However, the structure of these states is still
very different to those of a GOE system because $\avg{S^{inf}} \simeq \avg{S^{inf}_{GOE}}/2$. Only
when $\lambda \simeq 1$ one finds that $\avg{S^{inf}} \simeq \avg{S^{inf}_{GOE}}$, signaling that
the dynamics has become fully chaotic.

\section{Thermalization definitions and numerical results}
\label{sec:thermalization}

\subsection{Basic definitions}
\label{ssec:definitions}

We study the thermalization properties of finite fermionic systems with time-reversal and rotational
invariance. To be precise we consider $m$ fermions distributed in $n$ independent particle states,
interacting via the Hamiltonian \eref{eq:H(1+2)}, where $\varepsilon_k$ and $\BOK{p q}{V}{k l}$ are
independent real Gaussian random variables. These systems are usually called EGOE(1+2) ensembles. In
this work we use $\overline{\varepsilon_k} = k$, $\overline{(\epsilon_k - k)^2} = 1/2$,
$\overline{\BOK{p q}{V}{k l}} = 0$, $\overline{\left|\BOK{p q}{V}{k l}\right|^2} =
1+\delta_{(pq),(kl)}$, and the energy scale is such that $\rav{E} =0$ and $\rav{E^2}^{1/2}=1$,
regardless of the value of the interaction strength $\lambda$.  The results presented below have
been obtained by fully diagonalizing $60$ member EGOE(1+2) systems with $m=5,6$, $n=12-16$ and
$\lambda\in[0,1]$. The corresponding dimensions are given in Table \ref{tab:ensembles}.

\begin{table}
\begin{center}
\begin{tabular}{c|cccccccccc}\hline\hline
$m$ &    5    &    5    &    5    &   5    &   5    &    6  &    6  &   6   &    6  &   6    \\
$n$ &   12    &   13    &   14    &  15    &  16    &   12  &   13  &  14   &   15  &  16    \\
$d$ &  792    &  1287   &  2002   & 3003   & 5005   &  924  &  1716 & 3003  &  5005 & 8008  \\ \hline 
\end{tabular}
\caption{\label{tab:ensembles} Matrix dimension for EGOE systems with $m=5,6$ and $n=12-16$}
\end{center}
\end{table}

Let $\ket{\Psi(0)}$ be the initial state of the system, that we decompose as 

\be
\ket{\Psi(0)} = \sum_{\mu} C_{\mu} \Ket{\mu},
\label{ini_state}
\ee  
in the eigenstate basis $\lk \Ket{\mu}, \mu = 1,2,\dots d\rk$ of the Hamiltonian
\eref{eq:H(1+2)}. Then, given a certain observable $\Op$ we define:

\begin{itemize}
\item The instantaneous value $O(t) = \KOK{\Psi(t)}{\Op} = \tr{\Op \rho_{\Psi}(t)}$, where
  $\rho_{\Psi}(t) = \ket{\Psi(t)} \bra{\Psi(t)}$.
  
\item The time average $\tav{O(t)} = (2T)^{-1} \int_{t-T}^{t+T} O(\tau) \rmd\tau$. When $T
  \gg 1$, $\tav{O(t)} \xsim{T} \rav{\Op}_{eq} = \sum_{\mu} \lv C_{\mu} \rv^2 \DO{\mu}$, where
  $\DO{\mu}=\KOK{E_{\mu}}{\Op}$. The steady state average can also be written as $\rav{\Op}_{eq} =
  \tr{\Op \rho_{eq}}$, with $\rho_{eq} = \sum_{\mu} \lv C_{\mu} \rv^2 \Ket{\mu} \Bra{\mu}$.

\item The statistical average $\rav{\Op}_{stat} = \tr{\Op \rho_{stat}}$, where $\rho_{stat}$ is 
      the density operator corresponding to an appropriate statistical ensemble. 
\end{itemize}

We say that the system thermalizes if for any relevant observable $\Op$ and almost any state
$\ket{\Psi(t)}$, it is satisfied that 

\be 
\tav{O(t)} \xsim{T} \rav{\Op}_{eq} \approx \rav{\Op}_{stat}
\label{thermalization}
\ee

In practice, to asses whether Eq. \eref{thermalization} do approximately holds for a specific
observable, one can use the relative error

\be
\Delta_o = \abs{\dfrac{\delta_o}{\rav{\Op}_{stat}}} = \abs{\dfrac{\rav{\Op}_{eq}-\rav{\Op}_{stat}}{\rav{\Op}_{stat}}},
\label{DeltaO}
\ee
which is well suited to compare the degree of thermalization of different systems, or the
thermalization of a single system relative to distinct observables.

Let us consider that the system is prepared in a non-equilibrium state $\ket{\Psi(0)}$ for which the
energy is essentially constant. Hamiltonian eigenvectors are excluded as they are stationary states,
but we can assume that $\KOK{\Psi(0)}{H} = E_0$, and $\lb \KOK{\Psi(0)}{H^2-E_0^2}\rb^{1/2} < \Delta
E$, with $\Delta E$ sufficiently small compared to the energy spectrum span, but large enough to
contain many energy eigenstates. In such a case the microcanonical ensemble is the preferred
statistical ensemble. Denoting by $W$ the corresponding energy shell, i.e., $W = \{\Ket{\mu} ;
E_{\mu} \in [E_0-\Delta E,E_0+\Delta E]\}$, the microcanonical density operator is

$$\rho_{mc} = \frac{1}{d'}\sum_{\mu}{}^{'} \Ket{\mu} \Bra{\mu},$$
where $d'$ is the dimension of the subspace $W$ and the symbol
$\sum^{'}$ means that the sum is restricted to eigenstates belonging
to $W$. Thus, the corresponding microcanonical average is given by

$$\rav{\Op}_{mc} = \dfrac{1}{d'}\sum_{\mu}{}^{'} \DO{\mu}.$$

To examine whether the energy plays a role in the thermalization of the system we shall use three
different energy intervals with $\Delta_E=0.1$ and $E_0=0.0, 1.0$ and $1.7$. We denote the
corresponding energy shells by $W_1$, $W_2$ and $W_3$ respectively. Moreover, to see how the initial
conditions affect the process we let the system evolve from three different types of initial states,
defined as:

\begin{itemize}
\item $\ket{\Psi^{(1)}(0)} \propto P_{W} \ket{k_0}$, where $P_W$ is the projector onto $W$, and
      $\ket{k_0}$ is the mean-field state with energy $e = E_0$.

\item $\ket{\Psi^{(2)}(0)}  \propto \sum^{'}_{\mu} C_{\mu}\Ket{\mu}$, with the coefficients
      $C_{\mu}$ Gaussian random variables with mean zero and variance equal to one, i.e., 
      $C_{\mu} = \mathcal{G}(0,1)$.

\item $\ket{\Psi^{(3)}(0)} \propto \sum^{'}_{\mu} C_{\mu}\Ket{\mu}$, where the expansion
  coefficients are $C_{\mu} =  e^{-\alpha\lp\frac{E_{\mu}-E_0}{\Delta E}\rp^2 }\mathcal{G}(0,1)$.
\end{itemize}

The states $\Psi^{(2)}(0)$ and $\Psi^{(3)}(0)$ are random superpositions of the eigenstates
belonging to $W$, but due to the Gaussian factor the distribution of the $C_{\mu}$ coefficients is
wider for $\Psi^{(3)}(0)$. As we shall see below the distribution of the state amplitudes inside $W$
is one of the factors that may affect the thermalization process.

Before we turn to the main results, let us introduce the four types of generic observables for which
we study the validity of Eq. \eref{thermalization}:

\begin{itemize}
\item diagonal one-body operators $\Op{_d(1)} = \sum_{k} \theta_{k} a^{\dagger}_k a_k$,
\item one-body operators $\Op{(1)} = \sum_{k,l} \theta_{kl} a^{\dagger}_k a_l$,
\item two-body operators $\Op{(2)} = \sum_{k<l,p<q} \theta_{klpq} a^{\dagger}_k a^{\dagger}_l a_q a_p$,
\item strength function operators $\Op{_{sf}} = \Op{^{T}(1)}\Op{(1)}$,
\end{itemize}
where the parameters $\theta_{k}$, $\theta_{kl}$ and $\theta_{klpq}$ are taken as random variables.

\subsection{Numerical results}
\label{ssec:NResults}

\begin{figure}[h]
\begin{center}
\includegraphics[width=0.4\textwidth,height=0.4\textwidth,angle=-90]{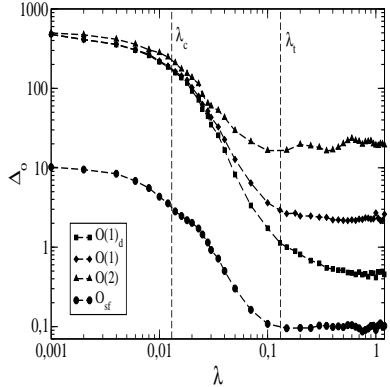} 
\vskip-3ex
\caption{Evolution with the interaction strength $\lambda$ of the averages $\avg{\Delta_{o_d(1)}}$
  (squares), $\avg{\Delta_{o(1)}}$ (diamonds), $\avg{\Delta_{o(2)}}$ (triangles), and
  $\avg{\Delta_{o_{sf}}}$ (dots), given in percent, for a $60$ member EGOE(1+2) with $(m,n)=(6,16)$
  initially prepared in a state $\Psi^{(1)}(0) \in W_1$.}
\label{fig:AvgDeltaOperators_6x16_60mt_st3E0}
\end{center}
\end{figure}

Let us consider the ensemble averages $\avg{\Delta_{o_d(1)}}$, $\avg{\Delta_{o(1)}}$,
$\avg{\Delta_{o(2)}}$, and $\avg{\Delta_{o_{sf}}}$ for a $60$ member EGOE with $(m,n)=(6,16)$.
Fig. \ref{fig:AvgDeltaOperators_6x16_60mt_st3E0} displays their evolution with the strength
$\lambda$ when the system is prepared at $t=0$ in a state $\Psi^{(1)}(0)$ with energy $E \simeq E_0
=0$ ( $\Psi^{(1)}(0) \in W_1$). Because of the large differences between the relative errors of
these operators and to properly visualize their evolution in the short interval $\lambda_c \le
\lambda \le \lambda_t$, a log-log scale is used. As for Fig. \ref{fig:AvgChaosMarkers_6x16_60mt} the
two vertical lines give the positions of $\lambda_c$ and $\lambda_t$. In all the cases
$\avg{\Delta_o}$ becomes smaller as the interaction strength increases up to $\lambda \approx
\lambda_t$. It is very important to realize that the transition from Poisson to GOE spectral
fluctuations, which is considered the most relevant signature of quantum chaos, occurs at $\lambda
\approx \lambda_c$ and does not modify this trend. On the contrary, for $\lambda > \lambda_t$ the
relative errors either remain essentially constant or the decreasing rate is much smaller.  Recall
that $\lambda_t$ defines a region where the three entropies $S^{ther}_E$, $S^{inf}_E$ and $S^{sp}_E$
take essentially the same values, and signals the point at which the wave functions start to become
very delocalized in the mean-field basis. Beyond $\lambda_t$, $\avg{\Delta_O}$ becomes clearly
smaller than one percent only for two operators, namely $\Op{_d(1)}$ and $\Op{_{sf}}$. Their errors
are $\avg{\Delta_{o_d(1)}} \approx 0.5$\% and $\avg{\Delta_{o_{sf}}} \approx 0.1$\%, respectively.
Thus, as long as the the system is prepared in an initial state $\Psi^{(1)}(0) \in W_1$ and $\lambda
> \lambda_t$, Eq. \eref{thermalization} approximately holds for the observables $\Op{_d(1)}$ and
$\Op{_{sf}}$ and then we can assert that the system thermalizes relative to these observables. This
is not the case of the observables $\Op{(1)}$ and $\Op{(2)}$. It is worth noting that perhaps the
main difference between $\Op{(1)}$, $\Op{(2)}$ in one hand, and $\Op{_d(1)}$, $\Op{_{sf}}$ in the
other, is that the latter have meaningful smoothed form for large $\lambda$, as given by spectral
distribution methods~\cite{Gomez_Kar:11,Kota:03,Kota_Haq:10}.

\begin{figure}[h]
\begin{center}
\includegraphics[width=0.4\textwidth,height=0.4\textwidth,angle=-90]{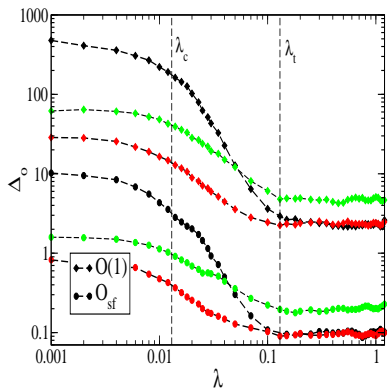} 
\vskip-3ex
\caption{Values of $\avg{\Delta_{o(1)}}$ (diamonds) and $\avg{\Delta_{o_{sf}}}$(dots), expressed in
  percent, as function of $\lambda$ in a $60$ member EGOE(1+2) with $(m,n)=(6,16)$, and initial
  conditions given by $\Psi^{(1)}(0)$ (black), $\Psi^{(2)}(0)$ (red), and $\Psi^{(3)}(0)$
  (green). In all cases the initial state belongs to the energy shell $W_1$.}
\label{fig:AvgDeltaOperators_6x16_60mt_st123E0}
\end{center}
\end{figure}

The results corresponding to other choices of $\Psi(0) \in W_1$ are shown in
Fig. \ref{fig:AvgDeltaOperators_6x16_60mt_st123E0}. For simplicity we have represented only the
results for $\Op{_{sf}}$ and $\Op{(1)}$. Curves in black, red and green correspond to
$\Psi^{(1)}(0)$, $\Psi^{(2)}(0)$ and $\Psi^{(3)}(0)$, respectively. In all cases the initial state
belongs to the energy shell $W_1$. We see that the choice of the initial conditions does not affect
the main trend: $\avg{\Delta_{o(1)}}$, and $\avg{\Delta_{o_{sf}}}$ diminish progressively as the
strength $\lambda$ is increased, and once $\lambda > \lambda_t$ their values remain essentially
constant. However, the precise values are quite different. When $\lambda> \lambda_t$, the initial
states $\Psi^{(1)}(0)$ and $\Psi^{(2)}(0)$ give rise to very similar results while the error
corresponding to $\Psi^{(3)}(0)$ is clearly larger.

\begin{figure}[h]
\begin{center}
\includegraphics[width=0.4\textwidth,height=0.4\textwidth,angle=-90]{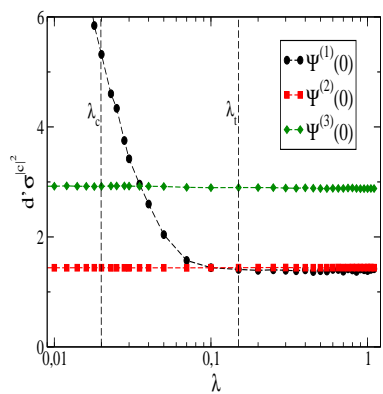} 
\vskip-3ex
\caption{Variation with the interaction strength $\lambda$ of the standard deviation $\sigma_{\lv
    C_{\mu} \rv^2}$ of the initial state $\Psi(0)$ components in the energy eigenbasis. The values
  of $\sigma_{\lv C_{\mu} \rv^2}$ have been multiplied by the dimension $d'$ of the energy shell. The
  results for $\Psi^{(1)}(0) \in W_1$ (black), $\Psi^{(2)}(0) \in W_1$ (red), and $\Psi^{(3)}(0) \in
  W_1$ (green) are plotted}
\label{fig:C2sigma_over_mean_6x14_60mts}
\end{center}
\end{figure}

To shed some light on the mechanisms leading to these results lets us note that the difference 
 
\be
\delta_0 = \rav{\Op}_{eq}-\rav{\Op}_{mc} = \sum_{\mu}{}^{'} \lp \lv C_{\mu} \rv^2 - \dfrac{1}{d'}\rp \DO{\mu},
\ee
becomes very small when: the distribution of the coefficients $C_{\mu}$ is nearly flat, i.e, $\lv
C_{\mu} \rv^2 \approx 1/d'$, or the matrix elements $\DO{\mu}$ almost do not fluctuate inside the
energy shell $W$, and therefore $\DO{\mu} \approx \rav{\Op}_{mc}$. This condition is known in the
literature as the eigenstate thermalization hypothesis (ETH), which has been conjectured to hold in
chaotic quantum systems~\cite{Rigol_Dunjko:08,Srednicki:94}. The standard deviation of the initial
state components in the energy eigenbasis, $\sigma_{\lv C_{\mu} \rv^2}$, is plotted in
Fig. \ref{fig:C2sigma_over_mean_6x14_60mts}. When $\lambda$ is very small $\sigma_{\lv C_{\mu}
  \rv^2}$ takes very different values in the three cases, being quite larger for $\Psi^{(1)}(0)$.
However, in this case $\sigma_{\lv C_{\mu} \rv^2}$ undergoes a sharp decreasing up to $\lambda
\simeq \lambda_t$, where its values are very similar to those of $\Psi^{(2)}(0)$ and half those of
$\Psi^{(3)}(0)$.

\begin{figure}[h]
\begin{center}
\includegraphics[width=0.4\textwidth,height=0.4\textwidth,angle=-90]{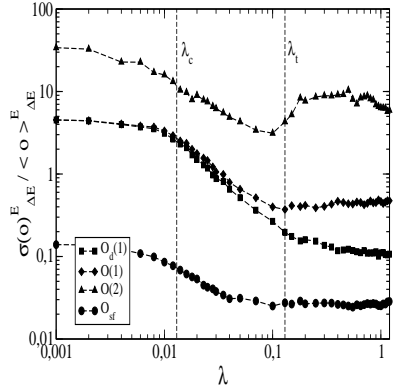}
\vskip-3ex
\caption{Evolution with $\lambda$ of the matrix elements $\KOK{E_{\mu}}{\Op}$ fluctuations in the
  central energy shell $W_1$, measured by the ratio $\kappa_{o} =\sigma_{o}^{mc} / \rav{\Op}_{mc}$
  (see text). Results for $\Op{_d(1)}$ (squares), $\Op{(1)}$ (diamonds), $\Op{(2)}$ (triangles) and
  $\Op{_{sf}}$ (circles).}
\label{fig:Mt20_1+2_6x16_BinOD_E0.ps}
\end{center}
\end{figure}

Defining the standard deviation of the expectation values $\DO{\mu}$ inside an energy shell 
$W$ as

\be
\sigma_{o}^{mc} = \lb \dfrac{1}{d'}\sum_{\mu}{}^{'} \KOK{E_{\mu}}{\Op-\rav{\Op}_{mc}}^2\rb^{1/2},
\ee
the ratio $\kappa_{o} = \sigma_{o}^{mc} / \rav{\Op}_{mc}$ provides us a measure of the
fluctuations of these matrix elements. Fig. \ref{fig:Mt20_1+2_6x16_BinOD_E0.ps} plots
$\kappa_{o}(\lambda)$ for a randomly selected member of the ensemble inside $W_1$.  Throughout the
interval $\lambda=0$ to $\lambda=1$ the ratio $\kappa_o$ is reduced sharply by a factor ranging
between 5 and 40, but even then we get $\kappa_{o} \gtrsim 1$ for operators like $\Op{(1)}$ and
$\Op{(2)}$. Remarkably, $\kappa_{o} \lesssim 0.1$ for $\Op{_d(1)}$ or $\Op{_{sf}}$, meaning
that the fluctuations of their expectation values are small compared to $\rav{\Op}_{mc}$ in the
energy shell $W_1$.

The extraordinary resemblance of Fig. \ref{fig:Mt20_1+2_6x16_BinOD_E0.ps} with
Figs. \ref{fig:AvgDeltaOperators_6x16_60mt_st3E0} and \ref{fig:AvgDeltaOperators_6x16_60mt_st123E0}
shows that the fluctuations of the eigenstate expectation values determine for which observables
thermalization occurs. In other words, the thermalization of the system is correlated with the
degree of compliance of the ETH. Moreover, given a certain observable the exact degree of
thermalization depends on the width of the initial state: the wider is the initial state the worse
thermalizes the system, a fact that is in agreement with some previous
claims~\cite{Santos_Rigol:11}.

\begin{figure}[h]
\begin{center}
\includegraphics[width=0.4\textwidth,height=0.4\textwidth,angle=-90]{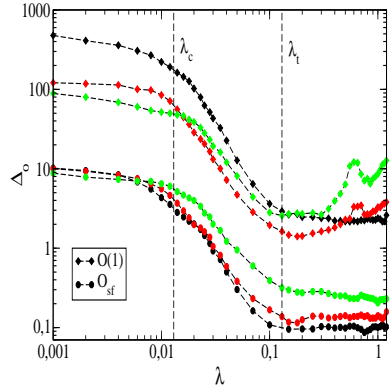}
\vskip-3ex
\caption{Behavior of the ensemble averages $\avg{\Delta_{o(1)}}$ (diamonds) and
  $\avg{\Delta_{o_{sf}}}$ (dots) as $\lambda$ varies, for three different initial state energies,
  $E_0 =0$ (black), $E_0 =1$ (red), and $E_0 =1.7$ (green).}
\label{fig:AvgDeltaOperators_6x16_60mt_st3E012}
\end{center}
\end{figure}

Thermalization may be affected by other factors, such as the proximity of $E_0$ to the edges of the
spectrum, and in particular to the ground-state energy.  In order to explore this question we have
compared the behavior of $\Delta_o$ in three energy shells $W_1$, $W_2$ and $W_3$, of width
$\Delta_E=0.1$ and $E_0=0.0, 1.0$ and $1.7$, respectively. Although the energies of the largest
eigenstates oscillate between $2.5$ and $3$ depending on the value of $\lambda$, it is very
difficult to get closer to the spectrum end. The reason is that outside the central interval
$(-2,2)$ the state density is so scarce that it is impossible to obtain a narrow energy shell which
contains a large number of states $d'$, but small compared to the dimension $d$ of the whole space.
We plot in Fig. \ref{fig:AvgDeltaOperators_6x16_60mt_st3E012} the variation with $\lambda$ of the
two averages $\avg{\Delta_{o_{sf}}}$ and $\avg{\Delta_{o(1)}}$, calculated for an EGOE system with
$(m,n)=(6,16)$, which is initially prepared in a state of type $\Psi^{(1)}(0)$. One sees immediately
that $\avg{\Delta_o}$ behaves very differently for these operators.  Once the thermalization region
($\lambda > \lambda_t$) has been reached, the relative error $\avg{\Delta_{o_{sf}}}$ grows as the
energy $E_0$ approaches the end of the energy spectrum. By contrast, the evolution of
$\avg{\Delta_{o(1)}}$ is more irregular, but still this error is clearly larger for $E_0 \approx
1.7$ and $\lambda \approx 1$. Therefore, the proximity of the initial state energy to the spectrum
edges inhibits the thermalization of the system.

\begin{figure}[h]
\begin{center}
\includegraphics[width=0.4\textwidth,height=0.4\textwidth,angle=-90]{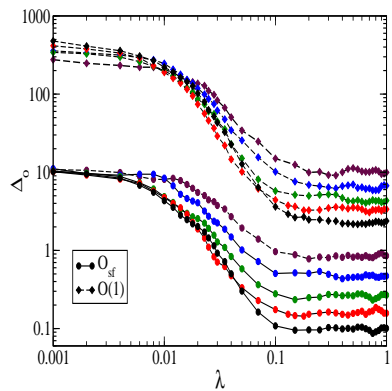}
\vskip-3ex
\caption{Variation with $\lambda$ of $\avg{\Delta_{o(1)}}$ (diamonds), and $\avg{\Delta_{o_{sf}}}$
(dots) in five ensembles with dimensions $d=924$ (maroon), $1716$   (blue), $3003$ (green), 
$5005$   (red), $8008$ (black), and initial conditions given by $\Psi^{(1)}(0) \in W_1$.}
\label{fig:AvgDeltaOperators_1+2_6xX_60mts_st3E0}
\end{center}
\end{figure}

Before closing this section we shall briefly consider whether the dimension of the $(m,n)$ Hilbert
space hinders or enhances thermalization. For illustration we consider $5$ systems with $(m,n) =
(6,12), (6,13), (6,14), (6,15)$ and $(6,16)$, with dimensions (see Table \ref{tab:ensembles})
$d=924, 1716, 3003, 5005$ and $8008$,
respectively. Fig. \ref{fig:AvgDeltaOperators_1+2_6xX_60mts_st3E0} shows the evolution of
$\avg{\Delta_o}$ for $\Op{_{sf}}$ and $\Op{(1)}$ as the dynamical regime changes from regularity at
$\lambda=0$ to full chaos for $\lambda \simeq 1$.  The growth rate of $d$ is very small, but it is
enough to induce a gentle decrease in $\avg{\Delta_o}$, while maintaining the same qualitative
trend. Similar results are obtained for the other two observables $\Op{_d(1)}$ and $\Op{(2)}$. If we
also take into account the analytical results of the next section we can conclude that the system
will always thermalize in the thermodynamic limit (the difference between microcanonical and
diagonal predictions decreases with system size), but there may be relevant differences for finite
dimensions, which is important since many physical systems of interest are mesoscopic.

\section{Analytical results for thermalization}
\label{sec:AResults}

Our goal now is to relate the relative error $\Delta_o$ with the fluctuation properties of 
the eigenstates and with the correlations generated by the observable $\Op$. To this end let us  
focus on the quantity

\be
\delta_0 = \rav{\Op}_{eq}-\rav{\Op}_{mc} = \tr{\Op (\rho_{eq}-\rho_{mc})} = \tr{\Op \Delta\rho}, 
\label{delta0}
\ee
with $\Delta\rho = \rho_{eq}-\rho_{mc}$. It can be considered as a random variable because
we model the physical system by means of an appropriate ensemble and thus the external
products $\Ket{\mu} \Bra{\mu}$ appearing in $\Delta\rho$ change from one member of the Hamiltonian
ensemble to another.  Moreover, being interested in ``typical'' properties we can introduce a
fictitious ensemble of initial states, $\ket{\Psi(0)} = \sum^{'}_{\mu} C_{\mu} \Ket{\mu}$, with
similar energies. Assuming that they are uniformly distributed in the unit sphere in $W$,
the fluctuation properties of the $C_{\mu}$ coefficients obey the Porter-Thomas (P-T) distribution
\cite{Brody_Flores:81} whenever $d'$ is large enough.  Thus

\be
\eav{\rho_{eq}} = \sum_{\mu}{}^{'} \eav{\lv C_{\mu}\rv^2} \Ket{\mu}\Bra{\mu} =
\dfrac{1}{d'}\sum_{\mu}{}^{'}\Ket{\mu}\Bra{\mu} =\rho_{mc},
\ee
where $\eav{\bullet}$ means averaging over the fictitious ensemble of initial states. Therefore
$\eav{\Delta\rho}=0$ implying that $\eav{\delta_0} = \avg{\eav{\delta_0}}=0$. Then,
$\avg{\eav{\delta^2_0}}$ is the variance of \eref{delta0} and we may agree on defining the
``typical'' value of $\Delta_o$ as

\be
\Delta^{typ}_o = \lb \dfrac{\avg{\eav{\delta^2_0}}}{\;\avg{\rav{\Op}^2}_{\!\!mc}} \rb^{1/2}.
\label{Delta_typ}
\ee

Introducing the shifted operator $\Op[Q] = \Op -aI$, with $a\in \mathbb{R}$, it is easy to see that
$\delta_0 = \delta_{Q}$. Moreover, a straightforward calculation gives

\be
\avg{\eav{\delta^2_0}} = \sum_{\mu,\nu}{}^{'}\, \avg{\DO[_Q]{\mu}\DO[_Q]{\nu}} 
\eav{ \lp \lv C_{\mu} \rv^2 - \dfrac{1}{d'}\rp \lp \lv C_{\nu} \rv^2 - \dfrac{1}{d'}\rp},
\ee
and taking into account the fluctuation properties of the $C_{\mu}$ coefficients we obtain 

\be
\avg{\eav{\delta^2_0}} = \dfrac{2}{d'^2} \sum_{\mu}{}^{'} \avg{\DO[^2_Q]{\mu}}.
\ee
Introducing this result in \eref{Delta_typ}, the value of $\Delta^{typ}_o$ is given by the
expression

\be
\Delta^{typ}_o = \lb 2 \dfrac{\sum{}_\mu^{'} \avg{\DO[^2_Q]{\mu}}}{\lp \sum{\!}_{\mu}^{'}\avg{\DO{\mu}}\rp^2} \rb^{1/2},
\label{Delta_typ2}
\ee
which can be further simplified because the averaged forms $\avg{\DO[^n_Q]{\mu}}$ are smooth
functions of the energy that do not change substantially inside the energy window $W$. Thus, we
arrive to

\be
\Delta^{typ}_o = \lb \dfrac{2}{d'} \dfrac{\avg{\DO[^2_Q]{0}}}{\lp \avg{\DO{0}}\rp^2} \rb^{1/2}.
\label{Delta_typ3}
\ee

\noindent
Since $a$ is a free parameter we can choose $a= \avg{\DO{0}}$, which gives rise to 

\be
\Delta^{typ}_o = \lb \dfrac{2}{d'} \dfrac{\avg{\DO[^2\!{}_o]{0}} - \lp \avg{\DO{0}}\rp^2}{\lp \avg{\DO{0}}\rp^2} \rb^{1/2}
             \!\!= \lb \dfrac{2}{d'} \dfrac{\avg{\sigma^2_o(E_0)}}{\lp   \avg{\DO{0}}\rp^2} \rb^{1/2}.
\label{Delta_typ4}
\ee

For strength function operators, like $\Op{_{sf}} = \Op^T\Op$, the value of $\Delta^{typ}_o$ can be
related to the NPC (also called IPR) in transition strengths originating from the central eigenstate
$\Ket{0}$.  Before we proceed it seems suitable to define the usual NPC for eigenstates and its
extension for transition strengths. The former is a measure of the eigenstate complexity in the
mean-field basis of Slater determinants $\ket{K}$ with $K \equiv {k_1,k_2,\dots,k_m}$ and
$k_i=1,2,\dots,n$. Expanding the eigenstates in this basis, i.e.,

\begin{equation}
\Ket{\mu} = \sum_{K=1}^d U_K(E_{\mu}) \ket{K},
\label{Eigvec_MeanField_Expansion}
\end{equation}
where the amplitudes $U_K(E_{\mu})$ satisfy that $\sum_{K=1}^d \lv U_K(E_{\mu}) \rv^2 =1$,
the number of principal components is defined as  

\be
\mbox{NPC}(E) = \lb \sum_K \lv U_K(E_{\mu}) \rv^4 \rb^{-1}.
\label{NPC_wf}
\ee

Note that NPC, as a function of the amplitudes $U$, attains its absolute minimum
$\mbox{NPC}_{\mbox{\tiny min}} = 1$ whenever the wave function is localized in a single basis state,
and its absolute maximum $\mbox{NPC}_{\mbox{\tiny max}} = d$ when the wave function is completely
delocalized and all the amplitudes are equal to $\lv U_K(E_{\mu}) \rv^2 = 1/d$.  Generally speaking
NPC gives the effective number of basis states that build up Hamiltonian eigenstates. Thus, it is
small for localized states, while for chaotic states, which are quite delocalized in the mean-field
basis, it takes values comparable but somewhat smaller than $d$ because system symmetries and
orthogonality prevent reaching its maximum. For instance, the average value for GOE is
$\avg{\mbox{NPC}}^{\mbox{\tiny GOE}} = d/3$.

Since the analysis of eigenvector amplitudes is largely equivalent to dealing with transition
strengths we can extend the previous discussion. Using the standard notation of spectral
distribution methods, let us introduce the so called locally normalized strength $\hat{R}$, and the
normalized stregth $\mathcal{R}$ generated by the action of the operator $\Op$ on a certain
eigenstate $\Ket{\mu}$ as

\begin{eqnarray} 
\hat{R}(E_{\nu},E_{\mu})     & = & \lb \avg{\lv \BOK{E_{\nu}}{\Op}{E_{\mu}} \rv^2} \rb^{-1} \lv \BOK{E_{\nu}}{\Op}{E_{\mu}} \rv^2,\\
\mathcal{R}(E_{\nu},E_{\mu}) & = & \lb M_0(E_{\mu}) \rb^{-1} \lv \BOK{E_{\nu}}{\Op}{E_{\mu}} \rv^2,
\end{eqnarray}
where $M_0(E_{\mu})= \KOK{E_{\mu}}{\Op^T\Op}$ is the total strength sum. In our notation
$M_0(E_{\mu}) = \DO[_{O_{sf}}]{E_{\mu}}$. Note that the initial and final spaces connected by $\Op$
do not need to be the same, but for simplicity we shall consider here a single space. If we normalize the
state vector $\Op \Ket{\mu}$ and expand it in the Hamiltonian eigenbasis

\be
\lb M_0(E_{\mu}) \rb^{-1} \Op \Ket{\mu} = \sum_{\nu=1}^d C_{\nu} \Ket{\nu} 
                                      = \sum_{\nu=1}^d \lb M_0(E_{\mu}) \rb^{-1} \BOK{E_{\nu}}{\Op}{E_{\mu}} \Ket{\nu},
\ee
then the NPC for the strength distribution generated by $\Op$ on the eigenstate $\Ket{\mu}$ is defined as

\be
\mbox{NPC}(E_{\mu}) = \lb \sum_{\nu} \lv C_{\nu} \rv^4 \rb^{-1} = \lb \sum_{\nu} \Big( \mathcal{R}(E_{\nu},E_{\mu}) \Big)^2 \rb^{-1}.
\label{NPC_sf}
\ee
Contrary to usual NPC in wavefunctions it only depends on the operator and the initial state, but
it is basis independent. The NPC for strength distributions gives the effective number of
eigenstates over which the strength generated by the action of the operator $\Op{}$ on a given
eigenstate is spread. For operators generating collective states (then $H$ and ${\mathcal O}$ are
highly correlated) NPC should be small, while for operators generating chaotic states it should take
large values.

Before turning back to the calculation of $\Delta^{typ}_{o_{sf}}$ we give an expression for the
smoothed NPC$_{E_{\mu}}$ that will be used below. If we assume that the fluctuations in the locally
renormalized strengths $ \hat{R}(E_{\nu},E_{\mu})$ follow the P-T distribution then $\avg{\hat{R}}=1$,
$\avg{\hat{R}^2}=3$, and 

\begin{eqnarray}
\avg{\mbox{NPC}}^{\mbox{\tiny EGOE}}_{E_{\mu}} & = \avg{\lb \sum_{\nu} \lp \mathcal{R}(E_{\nu},E_{\mu}) \rp^2 \rb^{-1}}
                                            = \lb 3 \sum_{\nu} \Big( \avg{\mathcal{R}(E_{\nu},E_{\mu})} \Big)^2 \rb^{-1} \nnb\\
                                          & = \lb 3 \sum_{\nu}\dfrac{\lp \avg{\BOK{E_\nu}{\Op}{E_0}^2}\rp^2}{\lp\avg{M_0(E_0)}\rp^2} \rb^{-1}
\label{Avg_NPC_sf}
\end{eqnarray}

\newpage

An EGOE(1+2) formula for NPC in transitions strengths and $\lambda$ not too far from $\lambda_t$ is given in
Ref. \cite{Kota_Sahu:98}. The typical error $\Delta^{typ}_{o_{sf}}$ can be written as

\be
\Delta^{typ}_{o_{sf}} = \lb \dfrac{2}{d'} \dfrac{\avg{\DO[^2\!{}_{o_{sf}}]{0}} - \lp \avg{\DO[_{o_{sf}}]{0}}\rp^2}{\lp \avg{\DO[_{o_{sf}}]{0}}\rp^2} \rb^{1/2} 
                    = \lb \dfrac{2}{d'} \dfrac{\avg{M^2_0(E_{0})} - \lp \avg{M_0(E_{0})}\rp^2}{\lp \avg{M_0(E_{0})}\rp^2} \rb^{1/2}, 
\ee
where 
\begin{eqnarray}
\avg{M^2_0(E_0)} &= \avg{\KOK{E_0}{\Op^T\Op}^2} = \avg{\lb \sum_{\nu} \BOK{E_\nu}{\Op}{E_0}  \rb^2} \nnb \\
                  &= \sum_{\nu} \avg{\BOK{E_\nu}{\Op}{E_0}^4} + \sum_{\nu\ne\mu} \avg{\BOK{E_\nu}{\Op}{E_0}^2}\;\avg{\BOK{E_\nu}{\Op}{E_0}^2}\\
                 &= \sum_{\nu} \lb \avg{\BOK{E_\nu}{\Op}{E_0}^4} - \lp \avg{\BOK{E_\nu}{\Op}{E_0}^2}\rp^2 \rb + \lp\avg{M_0(E_0)}\rp^2. \nnb
\end{eqnarray}
The step from the first to the second line in the previous expression follows from the independence
of the strengths due to the P-T assumption, and the third line is obtained by adding and subtracting
terms with $E_\nu=E_\mu$.  Using again that $\avg{\hat{R}^2}=3$, this result can be
simplified as

\be
\avg{M^2_0(E_0)} = 2 \sum_{\mu} \lp \avg{\BOK{E_\nu}{\Op}{E_0}^2}\rp^2  + \lp\avg{M_0(E_0)}\rp^2.
\label{M20_fin}
\ee

This expression was reported for the first time in \cite{Draayer_French:77} (see also
\cite{Kota:01}), and its predictions have been compared with shell model results in
\cite{Kota:01,Gomez_Kar:04} for the width of the strength sums fluctuations. Finally, 
inserting (\ref{M20_fin}) into (\ref{Avg_NPC_sf}), one gets

\be
\Delta^{typ}_{o_{sf}} = \lb \dfrac{4}{d'} \dfrac{\sum_{\mu} \lp \avg{\BOK{E_\nu}{\Op}{E_0}^2}\rp^2}{\lp\avg{M_0(E_0)}\rp^2} \rb^{1/2}
\!\!\!\!\!\!\!\!\!\!= \lb \dfrac{4}{3d'} \rb^{1/2}\!\!\!\! \lp \avg{\mbox{NPC}}^{\mbox{\tiny EGOE}}_{E_0} \rp^{-1/2},
\label{Delta_typ5}
\ee 
which establish a connection between the thermalization of the system, relative to an observable
$\Op{_{sf}} = \Op{^T(1)}\Op{(1)}$, and the value of the NPC for the transition strengths generated
by $\Op{(1)}$ acting on the eigenstate with energy $E_0$. An important outcome is that for chaotic
systems the NPC is expected to be large and hence these system will thermalize, while for regular
systems NPC has to be small and thus thermalization will be hindered. 

It is also possible to derive an expression for generic operators when $\lambda > \lambda_t$.  In
this case the eigenstates become more and more delocalized in the mean-field basis. Finally we reach
a regime were they behave as (quasi) random combinations of the basis states, the squared amplitudes
$\lv U_K \rv^2$ abide the P-T distribution and $S^{inf} = S^{inf}_{GOE}$ ($R^{inf} = 1$). In this
regime the expression \eref{Delta_typ4}, valid for a generic observable, and \eref{Delta_typ5} for
$\Op{_{sf}}$, take on a very simple form.  Using Eq. (\ref{Eigvec_MeanField_Expansion}) to
expand $\Ket{0}$ in the mean-field basis, one gets

\begin{eqnarray}
\avg{\DO{0}}  \overset{R^{inf} \rightarrow 1}{\Longrightarrow} & \sum_{K=1}^d \avg{U^2_K} \KOK{K}{\Op} = \dfrac{1}{d}\sum_{K=1}^d\KOK{K}{\Op} = \rav{\Op} ,\nnb\\
\avg{\DO{0}^2}\overset{R^{inf} \rightarrow 1}{\Longrightarrow} & \rav{\Op}^2 + \sum_{K=1}^d \avg{U^4_K-3U_K^2} \lv \KOK{K}{\Op} \rv ^2\\ 
                                                              &+ 2\sum_{K,K'=1}^d \avg{U_K^2}\;\avg{U_{K'}^2} \lv \BOK{K}{\Op}{K'} \rv^2 = \dfrac{2}{d}\rav{\Op^2} +\rav{\Op}^2, \nnb
\end{eqnarray}
 where the usual P-T formulas for the moments, $\avg{\lv U_K \rv^2}=1/d$ and $\avg{\lv U_K
\rv^4}=3/d^2$, have been applied.

\begin{figure}[h]
\begin{center}
\includegraphics[width=0.4\textwidth,height=0.4\textwidth,angle=-90]{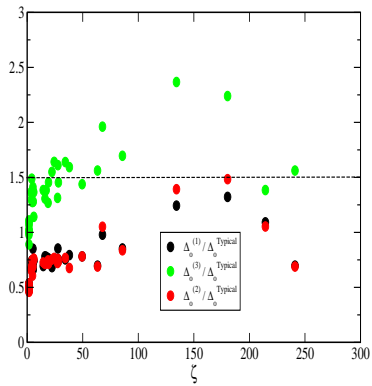}
\vskip-3ex
\caption{$\avg{\Delta_{o}} / \Delta^{typ}_{o}$ values for $\Op = \Op{_d(1)}, \Op{(1)}, \Op{(2)}$ and
  $\Op{_{sf}}$ in $60$ member EGOE(1+2) systems with $m=5,6$ and $n=12-16$, prepared at $t=0$ in
  initial states $\Psi(0) = \Psi^{(1)}(0)$ (black circles), $\Psi^{(2)}(0)$ (red squares) and
  $\Psi^{(3)}(0)$ (green diamonds), with energy $E \approx 0$.}
\label{fig:DeltaOperators_l=1_Typical_vs_exact2_60mts_st123E0}
\end{center}
\end{figure}

Gathering these expressions together with \eref{Delta_typ4} one finally arrives to  

\be
\Delta^{typ}_{o} = \lb \dfrac{4}{d d'} \dfrac{\rav{\Op^2}}{\rav{\Op}^2} \rb^{1/2},
\label{Delta_typ6}
\ee 
that links the value of $\Delta^{typ}_{o}$ with the dimension $d$ of the whole Hilbert space, the
dimension $d'$ of the microcanonical energy shell and the correlations generated by the operator
$\Op$, which are measured here by the ratio $\zeta = \lb \rav{\Op^2} / \rav{\Op}^2 \rb^{1/2}$.
Fig. \ref{fig:DeltaOperators_l=1_Typical_vs_exact2_60mts_st123E0} displays the values of
$\avg{\Delta_{o}}/ \Delta^{typ}_{o}$ calculated for the four observables $\Op{_d(1)}, \Op{(1)},
\Op{(2)}$ and $\Op{_{sf}}$ in EGOE(1+2) systems with $m=5,6$ and $n=12-16$. These are prepared in
three different initial states $\Psi(0) = \Psi^{(1)}(0), \Psi^{(2)}(0)$ and $\Psi^{(3)}(0)$, with
energy $E \approx 0$. It is clearly seen that $0 \le \avg{\Delta_{o}}/ \Delta^{typ}_{o} \le 3$,
confirming that $\Delta^{typ}_{o}$ really behaves as a standard deviation.

Using propagation formulas for $\rav{\Op{}}$ and $\rav{\Op{^2}}$ in terms of $m$ and $n$ one can
obtain particular expressions of $\Delta^{typ}_{o}$ for the operators considered in this paper. For
instance, following the results given in Appendix E of~\cite{Kota:01} we have
     
\be
\Delta^{typ}_{o_d(1)} \sim \lb \dfrac{4}{d d'} \lk 1 + \dfrac{1}{m}\dfrac{\rav{\theta^2}}{\rav{\theta}^2} \rk\rb^{1/2},
\label{Delta_typ7}
\ee 
for $O_d(1)$ in the so called dilute limit, i.e., $m,n\longrightarrow \infty, m/n \longrightarrow
0$.  Similar expressions can be derived for the other operators. 

An important outcome is the presence of the factor $(dd')^{-1/2}$. Since the energy window must be
sufficiently narrow so that the state density is constant inside, but wide enough to contain a large
number of states, we can assume that $d'= d / x$, where $x$ is essentially a fixed number. For
instance, in the energy shell $W_1$ we have $x \simeq 14$. This result suggest that $\Delta^{typ}_o$
is inversely proportional to the size of Hilbert space. However it should be noted that this trend
may be modified by sharp variations of the correlations measured by $\zeta$ (or
$\rav{\theta^2}/\rav{\theta}^2$) and by the fluctuations of the actual value $\Delta_o$ with regard
to $\Delta^{typ}_o$.

\begin{figure}[h]
\begin{center}
\includegraphics[width=0.4\textwidth,height=0.4\textwidth,angle=-90]{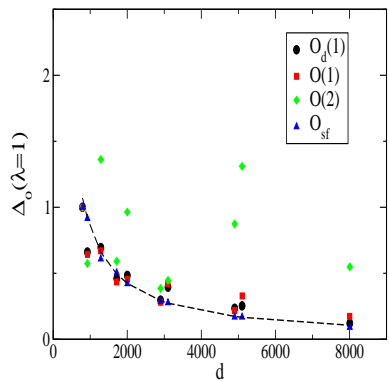}
\vskip-3ex

\caption{Evolution of the relative errors $\avg{\Delta_{o_d(1)}}$, $\avg{\Delta_{o(1)}}$,
  $\avg{\Delta_{o(2)}}$, and $\avg{\Delta_{o_{sf}}}$, calculated at $\lambda=1$, as the dimension of
  the Hilbert space increases. Their values have been rescaled so that $\avg{\Delta_o} = 1$ for
  $d=792$. The dotted line stands for $\Delta_o=a/d$, where the value $a\approx 848$ has been
  obtained by a least-squares fit to the data.}
\label{fig:DeltaOperators_l=1_vs_dimension_st3E0}
\end{center}
\end{figure}

Fig. \ref{fig:DeltaOperators_l=1_vs_dimension_st3E0} displays the evolution of $\avg{\Delta_o}$ at
$\lambda=1$ as the dimension of the Hilbert space increases. The values of $\avg{\Delta_o}$ have
been rescaled so that $\avg{\Delta_o} = 1$ for $d=792$. In most cases the values of $\avg{\Delta_o}$
fall very close to the $\Delta_o \propto d^{-1}$ law, represented by the dotted line.  The
fluctuations of the relative error $\avg{\Delta_{o(2)}}$ are clearly larger due to larger
oscillations of the corresponding $\zeta$ values. This fact is largely consistent with our previous
findings because one expects the correlations of two-body operators to be more complex than those of
one-body operators. In order to determine more precisely if the $d^{-1}$ is valid larger dimensions
would be required, but this is a formidable computational task. Nevertheless, these results suggest
that fully chaotic systems will thermalize relative to most observables in the thermodynamic limit.

\section{Conclusions}
\label{sec:Conclusions}

We have studied the thermalization process in isolated fermionic systems, described by EGOE(1+2)
ensembles. Our thermalization criterion relies on the ergodicity principle for the expectation of
observables, which is more general than other criteria previously used, like the representability of
occupancies by the Fermi-Dirac or Bose-Einstein distributions. Two-body random ensembles
are paradigmatic models to study quantum chaos and the dynamical transition from integrability to
chaos. It is well known that as the strength of the residual interaction is increased these systems
undergo a order-to-chaos transition, characterized by several chaos markers like $\lambda_c$, that
signals the onset of GOE spectral fluctuations, and $\lambda_t$ which defines a region where
different definitions of thermodynamic variables give essentially the same results, as it occurs for
infinite systems. For $\lambda > \lambda_t$ the Hamiltonian eigenstates become more and more
delocalized in the mean-field basis.  We have shown by means of exact diagonalizations that the
onset of Wigner spectral fluctuations is not sufficient to guarantee thermalization in finite
systems. Only if all chaos signatures are fulfilled, including the quasi complete delocalization of
eigenstates, we find that thermalization occurs for certain types of observables, such as (linear
combinations of) occupancies and strength function operators.

As stated by many authors, the eigenstate thermalization hypothesis seems to be the mechanism
responsible for thermalization, but in turn it holds only for certain observables provided that the
eigenstates behave as quasi-random superpositions of basis states. We have also analyzed the
influence of other factors, and found that the proximity of the initial state to the spectrum edges
hinders thermalization. On the contrary large Hilbert space dimensions and initial states with a
very small energy incertitude, i.e., with a very narrow distribution in the energy eigenbasis,
enhance the thermalization process. 

Analytical expressions linking the degree of thermalization for a given observable $\Op{}$ with
different properties of the system have been deduced. For instance, we have found that the typical
value of the relative error between the equilibrium and microcanonical averages, $\Delta^{typ}_o$,
is inversely proportional to the square root of the NPC (or IPR) for transition strengths generated
by $\Op{}$ acting on the middle of the microcanonical shell. An important outcome of this result is
that for chaotic systems the NPC is expected to be large and hence these system will thermalize,
while for regular systems NPC has to be small and thus thermalization will be hindered.

Similarly, we have found that when the eigenstates become fully delocalized, $\Delta^{typ}_0$ is
proportional to the square root of the correlations generated by the observable and inversely
proportional to the square root of the dimension $d$ of the whole Hilbert space times the dimension
$d'$ of the microcanonical energy shell .  Since $d' \propto d$, this result shows that
$\Delta^{typ}_0 \propto d^{-1}$ and suggests that fully chaotic systems will thermalize relative to
most observables in the thermodynamic limit.

In conclusion we have presented the first study of thermalization in the two-body random matrix
ensembles using the ergodicity principle for the expectation values of observables.

\section*{Acknowledgments}
The authors A. R. and J. R. are thankful to L. Mu\~noz and J. M. G. G\'omez for their collaboration
and enlightening discussions. V. K. B. K. and M. V. thank Navinder Singh for useful
discussions. This work is supported in part by Spanish Government grants for the research projects
FIS2006-12783-C03-02, FIS2009-11621-C02-01, CSPD-2007-00042-Ingenio2010, and by the Universidad
Complutense de Madrid grant UCM-910059. One of us, A. R., is supported by the spanish program JAE-Doc.


\begin{thebibliography}{xx}
\bibitem{omegaGOE}               It is also worth noting here that the actual value of the Brody parameter for GOE is not $\omega=1$, but $\omega = 0.965$.
\end{thebibliography}
\end{document}